\documentstyle[12pt]{article}
\input epsf.sty
\def\ZZZ{{\hbox{ Z\kern-1.6mm Z}}}

\newcommand{\AAA}{{\cal A}}
\newcommand{\GG}{{\cal G}}
\newcommand{\KK}{{\cal K}}

\newcommand{\FF}{{\cal F}}

\newcommand{\MM}{{\cal M}}

\newcommand{\OO}{{\cal O}}

\newcommand{\LL}{{\cal L}}

\newcommand{\square}{\Box}

\newcommand{\wt}{\widetilde}
\newcommand{\wh}{\widehat}
\newcommand{\wc}{\check}

\newcommand{\NN}{{\cal N}}

\newcommand{\SSS}{{\cal S}}

\newcommand{\be}{\begin{equation}}
\newcommand{\ee}{\end{equation}}
\newcommand{\ben}{\begin{eqnarray}\displaystyle}
\newcommand{\een}{\end{eqnarray}}
\newcommand{\refb}[1]{(\ref{#1})}
\newcommand{\p}{\partial}
\newcommand{\sectiono}[1]{\section{#1}\setcounter{equation}{0}}

\def\one{{\hbox{ 1\kern-.8mm l}}}
\def\zero{{\hbox{ 0\kern-1.5mm 0}}}
 
\begin{document}
{}~
{}~
\hfill\vbox{\hbox{hep-th/0411255}
}\break
 
\vskip .6cm
\begin{center}
{\Large \bf 
How Does a Fundamental String
Stretch its Horizon?
}

\end{center}

\vskip .6cm
\medskip

\vspace*{4.0ex}
 
\centerline{\large \rm
Ashoke Sen}
 
\vspace*{4.0ex}

\centerline{\large \it Harish-Chandra Research Institute}

\centerline{\large \it  Chhatnag Road, Jhusi,
Allahabad 211019, INDIA}
 
\centerline{E-mail: ashoke.sen@cern.ch,
sen@mri.ernet.in}
 
\vspace*{5.0ex}
 
\centerline{\bf Abstract} \bigskip

It has recently been shown that if we take into account a class of higher
derivative corrections to the effective action of heterotic string theory,
the entropy of the black hole solution representing elementary string
states correctly reproduces the statistical entropy computed from the
degeneracy of elementary string states. So far the form of the solution
has been analyzed at distance scales large and small compared to the
string scale. We analyze the solution that interpolates between these two
limits and point out a subtlety in constructing such a solution due to the
presence of higher derivative terms in the effective action. We also study
the T-duality transformation rules to relate the moduli fields of the
effective field theory to the physical compactification radius in the
presence of higher derivative corrections and use these results to find
the physical radius of compactification near the horizon of the black
hole. The radius approaches a finite value even though the corresponding
modulus field vanishes. Finally we discuss the non-leading contribution to
the black hole entropy due to space-time quantum corrections to the
effective action and the ambiguity involved in comparing this result to
the statistical entropy.

\vfill \eject
 
\baselineskip=18pt

\tableofcontents

\sectiono{Introduction and Summary} \label{sintro}

The idea that a very massive elementary string state should describe a
black hole is quite old\cite{thooft,9309145,9401070,9405117,9612146}.  
This leads one to wonder if the entropy associated with these black holes 
could be given a statistical interpretation as the degeneracy of the 
elementary string states of a given mass that the black hole represents. 
One of the problems in carrying out this exercise is that due to large 
renormalization effects it is often difficult to identify the class of 
elementary string states that represent a given black hole and vice 
versa\cite{9309145,9401070}.

One way to avoid this problem is to focus attention on BPS states for
which the renormalization effects are under control. In particular one can
consider heterotic string theory compactified on a torus and consider a
fundamental heterotic string wrapped along one of the circles of the
torus, carrying $w$ units of winding charge and $n$ units of momentum
along the same circle\cite{rdabh1,rdabh2}.
This describes a BPS state provided we do not
excite right-moving world-sheet oscillators. The degeneracy of such states
grow as $\exp(4\pi\sqrt{nw})$ for large $nw$, which suggests that we can
assign a statistical entropy of $4\pi\sqrt{nw}$ to these states. On the
other hand one can construct extremal BPS black hole solutions carrying
the same charge quantum numbers as these states. Thus one might hope that
the entropy of the black hole, computed using the Bekenstein-Hawking
formula, might reproduce the statistical entropy computed from the
degeneracy of the elementary string states.  Unfortunately the 
corresponding black hole has zero area of the event horizon and 
consequently the Bekenstein-Hawking entropy vanishes\cite{9504147}.

This however is not the end of the story. The black hole solution that 
gave vanishing entropy 
was constructed using 
tree level low 
energy effective action of the heterotic string theory where we ignore all 
terms containing more than two derivatives. However if we examine the 
solution carefully we discover that the Riemann curvature blows up at the 
horizon and hence the higher derivative terms cannot be ignored. One also 
finds that in the region where the curvature associated with the string 
metric is of order unity, the string 
coupling constant is small for large $nw$. Thus we expect that 
for large $nw$ the full solution will receive corrections from higher 
derivative tree level contribution to the effective action, but the effect 
of string loop corrections can be ignored.

Although we do not know the precise form of these higher derivative 
corrections, it was shown in \cite{9504147} using a simple scaling 
argument that any correction to the black hole entropy due to these tree 
level higher derivative terms must be of the form $a\sqrt{nw}$ where $a$ 
is a purely numerical constant. This clearly agrees with the form of the 
statistical entropy. However the coefficient $a$ could not be calculated 
at that time.

Recently in a beautiful paper\cite{0409148} Dabholkar computed the
coefficient $a$ by including in the effective action a class of higher
derivative terms. These terms arise from the 
supersymmetric generalization
of the curvature squared 
term which is known to be present in the tree level
effective action of heterotic string theory\cite{rzwiebach,9610237}.
Following earlier 
work\cite{9601029,9812082,9904005,9906094,9910179,0009234},
ref.\cite{0409148} showed that the black hole solution is modified near
the horizon in a way that precisely reproduces the correct value $4\pi$
for the coefficient $a$. In arriving at this result one needs to take into
account not only the change in the area of the event horizon (which only
accounts for half of the entropy) but also a suitable modification of the
Bekenstein-Hawking entropy formula in the presence of the higher
derivative terms\cite{9307038,9403028,9502009}.\footnote{In this context
we note that the scaling argument of \cite{9504147} holds even in the
presence of such corrections to the entropy formula. The only assumption
required for this argument is that if we change the overall normalization 
of the action by a
constant, then the entropy associated with a given black hole solution
gets multiplied by the same constant. This will be reviewed in some detail
in section \ref{sreview}.} The key assumption behind this construction
is that the solution close to the horizon has maximal supersymmetry.

The analysis of \cite{0409148,0410076} gives the form of 
the solution only very close to the horizon, at distance scale much 
smaller than the string scale. 
On the other hand the near horizon 
solution\footnote{`Near horizon' here refers to distance scale small 
compared to the mass of the black hole.}  
based on the low energy limit of the effective field theory, described in 
\cite{9504147}, is expected to be valid only at a distance scale large 
compared to the string scale where higher derivative terms can be 
ignored.
Thus an important question that arises is: is there a smooth solution that
interpolates between these two limits? It turns out that the relevant
equation that needs to be analyzed is a second order non-linear
differential equation and hence although both the near horizon and the
large distance solutions satisfy this equation, it is not obvious that there
is a solution that interpolates between the two limiting solutions.
One of the goals 
of this paper will be to analyze this issue. Numerical analysis of the 
differential equation
indicates that if we begin with the near horizon solution and let it
evolve according to the equations of motion, the solution does not approach
the expected form at large radius, but oscillates about this form. Naively 
this would indicate that the solution does not approach the desired limit 
at large radius. However we argue that the supergravity description uses a 
choice of fields whose propagators have additional poles besides those 
implied by string theory, and once we make the correct choice of fields by 
using an appropriate field redefinition, these oscillations disappear 
and 
the solution approaches the correct asymptotic form at large 
radius.\footnote{Although our discussion will focus on the case of
two charge black hole representing elementary string states, a similar
subtlety is expected to arise for the three charge black hole which has
a finite area of the event horizon at the leading order.}

The form of the solution obtained in \cite{0409148} indicates that
the modulus field associated with the radius of the circle along which
the fundamental string is wrapped vanishes at the horizon. 
Naively this would imply that the radius of this circle vanishes
at the horizon. However, by analyzing the T-duality transformation
laws of various fields we show that the relationship between the physical
radius of the circle and the modulus field is modified in such a way
that the physical radius approaches a constant at the horizon even though
the associated modulus field vanishes.

Although for large charges the string coupling at the horizon is small
and hence we can ignore the effect of space-time
quantum corrections, 
ref.\cite{0409148} analyzes the non-leading contribution to the
entropy due to these quantum corrections. We reanalyze these effects and 
show
that if we define the statistical entropy as the logarithm of the 
degeneracy of states of the elementary string, then the geometric entropy
of the black hole fails to reproduce correctly the coefficient of the
term proportinal to $\ln(nw)$
in the expression for the statistical entropy. One should however keep
in mind that there are alternative definitions of the statistical 
entropy in terms of other ensembles, {\it e.g.} grand 
canonical ensemble, where we first introduce a grand canonical
partition function as a function of the chemical potential conjugate
to various charges, and then compute the entropy from this partition
function using the usual 
thermodynamic relations. These two definitions of entropy differ from
each other beyond the leading term, and it is not {\it a priori} 
clear as to which
definition of entropy should be compared to the geometric entropy of the
black hole. We show that one such definition of statistical entropy
agrees with the geometric entropy of the black hole beyond the leading 
order approximation.

The paper is organised as follows. We work in the 
$\alpha'=16$ unit as in \cite{9504147,9402002}. 
In section \ref{sreview} we
review the arguments of \cite{9504147} showing that the black hole entropy
has the correct dependence on various parameters up to an overall
numerical constant, and also review the recent results of
\cite{0409148,0410076}. In section \ref{scomplete} 
we construct the complete
near horizon solution, study the T-duality transformation rules of
various fields to determine the relation between the moduli fields and the
physical radius of compactification, and discuss the
effect of quantum corrections on the black hole entropy.
We end in section \ref{scomments} with some 
comments on possible generalizations and open issues. 

Possible importance of field redefinition in string theory
(or equivalently 
renormalization scheme dependence in two dimensional field
theory) in obtaining non-singular solution describing
a fundamental string has been discussed earlier in 
\cite{9504147,9509050}. Modification of black hole solutions
and T-duality rules due to higher derivative corrections to
the string effective action have been discussed earlier in
\cite{0408200} in a different context.

\sectiono{Supergravity Solution for Two Charge Black Holes 
and its Near
Horizon Limit} \label{sreview}

Although the analysis of \cite{0409148} is able to produce the 
complete formula for the geometric
entropy of the black holes describing
elementary string states, it relies on the assumption that the
contribution to the geometric entropy comes only from certain 
higher derivative terms
in the effective action. In contrast, the scaling argument of
\cite{9504147} does not rely on any such assumption, and hence
is still
of interest.
In this section we shall
first review the scaling argument
of ref.\cite{9504147}, and then briefly recall the results of
\cite{0409148}.

In
\cite{9504147} we
analyzed the most general electrically charged extremal 
black hole solution in heterotic string theory 
compactified on $T^6$. In order to keep our discussion simple, we 
shall here consider only a special class of black hole solutions 
representing a heterotic string wound on a circle. For this purpose we 
take heterotic string theory compactified on $T^5\times S^1$,
$T^5$ being an arbitrary five-torus and $S^1$ being a circle of 
coordinate radius 
$\sqrt{\alpha'}=4$. Let us denote by 
$x^\mu$ ($0\le\mu\le 3$) the non-compact directions and by $x^4$ the 
coordinate 
along $S^1$. As in \cite{9402002} we shall denote by $G^{(10)}_{MN}$, 
$B^{(10)}_{MN}$ and $\Phi^{(10)}$ the ten dimensional string 
metric, anti-symmetric tensor field and dilaton
respectively.
For 
the description of the black hole solution under study we shall only need 
to consider non-trivial configurations of the fields $G^{(10)}_{\mu\nu}$, 
$B^{(10)}_{\mu\nu}$, 
$G^{(10)}_{4\mu}$, $G^{(10)}_{44}$, $B^{(10)}_{4\mu}$ and $\Phi^{(10)}$. 
We freeze all other field components to trivial background values, and 
define:\footnote{Our convention for normalization 
of the dilaton is the 
same as that
in \cite{9504147,9402002}, \i.e. $e^{\Phi}$ represents the effective 
closed 
string coupling constant.}
 \ben \label{e3}
 && \Phi = \Phi^{(10)} - {1\over 2} \, \ln (G^{(10)}_{44})\, , \qquad 
S=e^{-\Phi}\, , \qquad 
 T = \sqrt{G^{(10)}_{44}}\, , \nonumber \\
&& G_{\mu\nu} = G^{(10)}_{\mu\nu} - (G^{(10)}_{44})^{-1} \, 
G^{(10)}_{4\mu} 
\, G^{(10)}_{4\nu}\, , \qquad 
g_{\mu\nu} = e^{-\Phi} \, G_{\mu\nu}\, ,\nonumber \\
&& A^{(1)}_\mu = {1\over 2} (G^{(10)}_{44})^{-1} \, G^{(10)}_{4\mu}\, , 
\qquad
A^{(2)}_\mu = {1\over 2} B^{(10)}_{4\mu}\, , \nonumber \\
&& B_{\mu\nu} = B^{(10)}_{\mu\nu} - 2(A^{(1)}_\mu A^{(2)}_\nu - 
A^{(1)}_\nu A^{(2)}_\mu)
\, .
 \een
The low energy effective action involving these fields is then given 
by\cite{9207016,9402002}
\ben \label{e4}
\SSS &=& {1\over 32\pi} \int d^4 x \, \sqrt{-\det g} \, \bigg[ R - {1\over 
2 
S^2} \, g^{\mu\nu} \, \p_\mu S \p_\nu S -  {1\over
T^2} \, g^{\mu\nu} \, \p_\mu T \p_\nu T 
\nonumber \\
&&- {1\over 12} S^2 g^{\mu\mu'} 
g^{\nu\nu'} g^{\rho\rho'} H_{\mu\nu\rho} H_{\mu'\nu'\rho'} - S T^2 \, 
g^{\mu\nu} \, g^{\mu'\nu'} \, F^{(1)}_{\mu\mu'} 
F^{(1)}_{\nu\nu'} - ST^{-2} \,
g^{\mu\nu} \, g^{\mu'\nu'} \, F^{(2)}_{\mu\mu'}
F^{(2)}_{\nu\nu'}\bigg] \, , \nonumber \\
\een
where
\ben \label{e4a}
&& F^{(a)}_{\mu\nu} = \p_\mu A^{(a)}_\nu - \p_\nu A^{(a)}_\mu\, , \quad 
a=1,2\, , \nonumber \\
&& H_{\mu\nu\rho} = \left[ \p_\mu B_{\nu\rho} + 2 \left( A_\mu^{(1)} 
F^{(2)}_{\nu\rho} + A_\mu^{(2)}
F^{(1)}_{\nu\rho}\right) \right] + \hbox{cyclic permutations of $\mu$, 
$\nu$, $\rho$}\, .
\een
In this normalization convention the 
Newton's constant is given by
\be \label{e5.0h}
G_N = 2\, .
\ee
Also for $H_{\mu\nu\rho}=0$ the $S$- and $T$-duality transformations
take the form\cite{9402002}:
\be \label{e4b}
S\to {1\over S}, \qquad F^{(1)}_{\mu\nu} \to - S\, 
T^{-2} \wt F^{(2)}_{\mu\nu},
\qquad F^{(2)}_{\mu\nu} \to - S\, T^2 \wt F^{(1)}_{\mu\nu}\, ,
\ee
and
\be \label{e4c}
T\to {1\over T}\, , \qquad F^{(1)}_{\mu\nu}\to F^{(2)}_{\mu\nu}, \qquad
F^{(2)}_{\mu\nu}\to F^{(1)}_{\mu\nu}\, ,
\ee
respectively.
$\wt F^{(a)}_{\mu\nu}$ denotes the Hodge dual of $F^{(a)}_{\mu\nu}$ 
with respect to the canonical metric $g_{\mu\nu}$.

We now consider an heterotic string wound $w$ times along 
the circle $S^1$ labelled by $x^4$ and 
carrying $n$ units of momentum along the same circle. 
Suppose further that 
asymptotically the four dimensional string coupling takes value $g$ and 
the radius of $S^1$ measured in the string metric takes value $R$. 
In our normalization convention this 
imposes the asymptotic conditions:
\ben \label{e5}
&& g_{\mu\nu} \to \eta_{\mu\nu} \nonumber \\
&& S\to g^{-2}, \qquad T\to R/4\, , \nonumber \\
&& F^{(1)}_{\rho t} \to 16\,  g^2 \, {n\over R^2} \, {1\over \rho^2}\, , \qquad
F^{(2)}_{\rho t} \to {1\over 16}\, g^2 \, w \, R^2 \, {1\over \rho^2}\, , 
\een
where $\rho$ is the radial distance from the black hole measured in the 
canonical metric $g_{\mu\nu}$.
An extremal black hole solution satisfying these 
asymptotic conditions can be read out from the general class of extremal 
black hole 
solutions constructed in \cite{9411187,9504147} (see also
\cite{rbreit})
and 
takes the form\footnote{In 
using the results of \cite{9504147} we should note
that appropriate components of
the right and the left-handed gauge fields given there 
correspond to ${1\over \sqrt 2} (A^{(1)}_\mu\pm 
A^{(2)}_\mu)$ of the present paper, and an appropriate $2\times 2$ block 
of the matrix $M$ given in \cite{9504147} can be identified to
the matrix ${1\over 2}\pmatrix{T^2+T^{-2} & T^{-2}-T^{2}\cr T^{-2} - 
T^{2} & 
T^2+T^{-2}}$ in the convention of the present paper. In order to 
produce the solution \refb{e6} from the one given in 
\cite{9504147}, we take $Q_R$, $Q_L$ of \cite{9504147} to be
$2\sqrt 2 g^2 (n/R \pm w R/16)$ and then rescale 
the fields $T$, $A^{(1)}_\mu$ and $A^{(2)}_\mu$ by $R/4$, 
$4/R$ and $R/4$ respectively. The latter operation
is a symmetry of the effective action
\refb{e4}, and is needed in order to produce a solution
for which 
the asymptotic value of $G^{(10)}_{44}$ is $R^2/16$ 
so that the asymptotic
radius of $S^1$ is $R$.} 
\ben \label{e6}
ds_c^2 &\equiv& g_{\mu\nu} dx^\mu dx^\nu = - (F(\rho))^{-1/2} \rho dt^2 + 
(F(\rho))^{1/2} \, \rho^{-1} d\vec{x}^2 \, , \qquad \rho^2 = \vec x^2\, , 
\nonumber \\
S &=& g^{-2}  \, (F(\rho))^{1/2} \, \rho^{-1} \, , \nonumber \\
F(\rho)&=& (\rho + g w R/2) (\rho+8g n R^{-1}) \, , \nonumber \\
T &=& {1\over 4}\, 
R \, \sqrt{ (\rho+8g n R^{-1}) / (\rho + g w R/2)} \, , \nonumber \\
F^{(1)}_{\rho t}  &=& {16 g^2 R^{-2} n \over (\rho+8g n R^{-1})^2} \, , 
\nonumber \\
F^{(2)}_{\rho t} &=& {1\over 16}\, 
{g^2 w R^2 \over (\rho + g w R/2)^2} \, , \nonumber 
\\
H_{\mu\nu\rho} &=& 0\, .
\een
$ds_c$ denote the line element measured in the canonical metric 
$g_{\mu\nu}$. The line element $ds_{string}$ measured in the string metric 
$G_{\mu\nu}$ is given by:
\be \label{e7}
ds_{string}^2 \equiv G_{\mu\nu} dx^\mu dx^\nu = S^{-1} ds_c^2 
= -g^2 \, \rho^2 \, (F(\rho))^{-1} dt^2 + g^2 \, d\vec{x}^2 \, .
\ee

The (singular) horizon for this solution is located at $\rho=0$. The near 
horizon region is defined as
\be \label{enear}
\rho << 8g n R^{-1}, g w R/2\, .
\ee
In this region
the solution takes the form:
\ben \label{e8}
ds_{string}^2 &=& -{\rho^2 \over 4 n w} \, dt^2 + g^2 \, d\vec{x}^2 
\, , 
\nonumber \\
S &=& {2 \sqrt{nw}\over g\rho}\, , \nonumber \\
T &=& \sqrt{n\over w} \, , \nonumber \\
F^{(1)}_{\rho t}  &=& {1\over 4n} \, , \nonumber 
\\
F^{(2)}_{\rho t}  &=& {1\over 4w} \, , 
\nonumber
\\
ds_c^2 &=& -{\rho\over 2 \, g \, \sqrt{nw}} dt^2 + {2 \, g \, 
\sqrt{nw}\over \rho} \, d\vec x^2\, .
\een
We now introduce rescaled coordinates:
\be \label{e9}
\vec y = g \, \vec x, \qquad r = \sqrt{\vec y^2} = g \, \rho\, , \qquad 
\tau = g^{-1} t / \sqrt{nw}\, .
\ee
In this coordinate system the solution near the horizon takes the form:
\ben \label{e10}
ds_{string}^2 &=& -{r^2 \over 4} \, d\tau^2 + \, d\vec{y}^2 
\, , \qquad r^2 = \vec y^2\, ,
\nonumber \\
S &=& {2 \sqrt{nw}\over r}\, , \nonumber \\
T &=& \sqrt{n\over w} \, , \nonumber \\
F^{(1)}_{r\tau} &=& {1\over 4} \, \sqrt{w\over n}\, , \nonumber 
\\
F^{(2)}_{r \tau} &=& {1\over 4} \, \sqrt{n\over w} \, .
\nonumber
\\
\een
Notice that in this new coordinate system the solution near the horizon is 
determined completely by the charge quantum numbers $n$ and $w$ and is 
independent of the asymptotic value of the moduli $g$ and $R$. This is an 
example of the attractor mechanism for supersymmetric black 
holes\cite{9508072,9602111,9602136}.

We now note that the 
tree level low energy effective action involving charge neutral fields is 
invariant under a rescaling of the form:
\be \label{e11}
G^{(10)}_{44} \to e^{2\beta} G^{(10)}_{44}, \qquad G^{(10)}_{4\mu}\to 
e^{\beta} G^{(10)}_{4\mu}, \qquad B^{(10)}_{4\mu}\to
e^{\beta} B^{(10)}_{4\mu}\, ,
\ee
keeping the four dimensional dilaton $\Phi$ fixed.
Physically this corresponds to a rescaling of
the compactification radius by $e^\beta$.
Clearly the full string theory is 
sensitive to the radius of compactification
and is not invarinat under this 
transformation. However the tree level effective action involving charge 
neutral fields, which are involved in the construction of the black hole 
solution, is not sensitive to the compactification radius, and the action 
as well as all the quantities ({\it e.g.} the black hole entropy) computed 
from the effective action will be unchanged under this rescaling. In terms 
of the four dimensional fields defined in \refb{e3} this amount 
to:\footnote{This is a special case of the O(6,22;R) transformation that 
was used in \cite{9504147} to bring the near horizon limit of
a general black hole solution into the 
universal 
form.}
 \be 
\label{e12}
T\to e^\beta T, \qquad A^{(1)}_\mu \to e^{-\beta}  A^{(1)}_\mu, \qquad 
A^{(2)}_\mu \to e^{\beta}  A^{(2)}_\mu \, .
\ee
Choosing $e^\beta=\sqrt{w/n}$ we can map the near 
horizon solution \refb{e10} 
to:\footnote{We would 
like to emphasize that the checked and hatted solutions discussed in this 
section are related to the original solution \refb{e8} by transformations 
which are exact symmetries of the equations of motion of tree level string 
theory, but are not exact symmetries of the full string theory.} 
\ben \label{e13}
\wc{ds}_{string}^2 &=& -{r^2 \over 4} \, d\tau^2 + \, d\vec{y}^2
\, , \qquad r^2 = \vec y^2\, ,
\nonumber \\
\wc S &=& {2 \sqrt{nw}\over r}\, , \nonumber \\
\wc T &=& 1 \, , \nonumber \\
\wc F^{(1)}_{r\tau} &=& {1\over 4} \, , \nonumber
\\
\wc F^{(2)}_{r \tau} &=& {1\over 4} \, .
\nonumber
\\
\een
We now note that except for the overall multiplicative factor of 
$\sqrt{nw}$ in the expression for $\wc S$, the solution has no dependence 
on 
any parameter and is completely universal. We also note that the area of 
the event horizon, measured in the canonical metric 
$g_{\mu\nu}=SG_{\mu\nu}$, is given by:
\be \label{e14}
A_H = 4\pi r^2 \wc S|_{r=0} = 8\pi \sqrt{nw} \, r|_{r=0} = 0\, .
\ee
Thus the area of the event horizon vanishes. As a result the black hole 
entropy also vanishes to this approximation.

Before we proceed we would like to make the following observations:
\begin{itemize}
\item \refb{e13} is an exact solution of the classical low energy
supergravity equations of motion. This follows from the fact that \refb{e6}
is a solution of these
equations for all $n$ and $w$, and \refb{e13}
is obtained from this solution by taking the limit $n,w\to\infty$ and
carrying out operations which are exact symmetries of the 
classical low energy 
supergravity equations of motion.
\item For $r>>1$ the higher derivative corrections to the solution
\refb{e13}
are small and we expect the solution of the complete classical 
equations of
motion of string theory to be approximated
by \refb{e13} in this limit. This can be seen by introducing a new
coordinate $\eta$ via the relation 
$\tau=2\eta / r$, and writing the solution as
\ben \label{etrssol}
\wc {ds}_{string}^2 &=&   -d\eta^2 + \, d\vec{y}^2 +2\, {\eta\over r} 
d\eta dr
- {\eta^2\over r^2} dr^2
\, , \qquad r^2 = \vec y^2\, ,
\nonumber \\
\p_r \wc S / \wc S&=&  -1/r \, , \nonumber \\
\wc T &=& 1 \, , \nonumber \\
\wc F^{(1)}_{r\eta} &=& {1\over 2 r} \, , \nonumber
\\
\wc F^{(2)}_{r \eta} &=& {1\over 2 r} \, .
\nonumber
\\
\een
Thus we see that for fixed $\eta$, the metric approaches flat metric and all
other fields become trivial for large $r$. Thus we expect the corrections
due to higher derivative terms to be small. In fact from the structure
of \refb{etrssol} it is clear that for fixed $\eta$ each derivative with
respect to $r$ brings down a factor of $1/r$ and hence
the effect of the four derivative terms
in the action
is suppressed by a factor of $1/r^2$ relative to the two derivative terms.
Thus we expect that the modification of the solution \refb{e13} due to the
higher derivative terms will be of order $1/r^2$ relative to the leading
term. This observation will be useful for our analysis later.

\end{itemize}

Let us now consider the effect of various corrections to the effective 
action\cite{9504147}. First of all we see that $S\to \infty$ as $r\to 0$ 
and even 
for $r\sim 
1$, $S$ is of order $\sqrt{nw}$ which is large for large $n$ and $w$. 
Since $S$ measures the inverse of the string coupling we conclude that 
stringy quantum corrections can be ignored for large $n$ and 
$w$\cite{9504147}. On the other hand since various curvatures are of order 
unity for $r\sim 1$ we expect that the tree level higher derivative terms 
will affect the solution and the entropy. To study the general form of 
these corrections, we recall that the complete tree level effective action 
of the 
heterotic string theory in the subsector under study has the form:
\be \label{e15}
\SSS = \int d^4 x \, \sqrt{-\det G} \, S \, \LL(G_{\mu\nu}, 
B_{\mu\nu}, T, A_\mu^{(1)}, A_\mu^{(2)}, \p_\mu S / S)\, .
\ee
Note in particular that under multiplication of $S$ by a constant, the 
action gets multiplied by the same constant. This shows that given any 
solution of the full equations of motion derived from the action 
\refb{e15}, we can get another solution 
by multiplying $S$ by an arbitrary 
constant, leaving the rest of the fields unchanged. Thus in order to
study possible corrections to the solution \refb{e14} due to 
the higher derivative terms in the action \refb{e15}, we could 
first find corrections to a different solution 
\renewcommand{\wh}{\hat}
\ben \label{e16}
\wh{ds}_{string}^2 &=& -{r^2 \over 4} \, d\tau^2 + \, d\vec{y}^2
\, , \qquad r^2 = \vec y^2\, ,
\nonumber \\
\wh S &=& {2 \over r}\, , \nonumber \\
\wh T &=& 1 \, , \nonumber \\
\wh F^{(1)}_{r\tau} &=& {1\over 4} \, , \nonumber
\\
\wh F^{(2)}_{r \tau} &=& {1\over 4} \, ,
\nonumber
\\
\een
and then multiply the $\wh S$ 
for the resulting solution by $\sqrt{nw}$ to 
find the correction to \refb{e13}. Since \refb{e16} has a completely 
universal form without any parameter. and since furthermore the action 
\refb{e15} is also completely universal, it is clear that the higher 
derivative terms in \refb{e15} will change \refb{e16} to a universal form:
\ben \label{e17}
\wh{ds}_{string}^2 &=& - {f_1(r)\over f_3(r)} \, d\tau^2 + 
{f_2(r)\over f_3(r)} \, d\vec{y}^2
\, , \qquad r^2 = \vec y^2\, ,
\nonumber \\
\wh S &=& f_3(r)\, , \nonumber \\
\wh T &=&  f_4(r)\, , \nonumber \\
\wh F^{(1)}_{r\tau} &=& f_5(r) \, , \nonumber
\\
\wh F^{(2)}_{r \tau} &=& f_6(r) \, ,
\nonumber
\\
\een
where $f_1(r),\ldots f_6(r)$ are a set of universal functions.
This particular parametrization has been chosen for later
convenience. For large 
$r$ these functions must agree with the solution \refb{e16}. This gives
\be \label{e18}
f_1(r) \simeq 
{r \over 2} \, , \quad f_2(r) \simeq {2\over r}\, , \quad
f_3(r) \simeq {2 \over r}\, , \quad f_4(r) \simeq 1, \quad f_5(r) \simeq 
{1\over 4} \, , \quad
f_6(r) \simeq {1\over 4} \, .
\ee
The higher derivative corrections to \refb{e13} is now generated by 
multiplying $S$ in \refb{e17} by a factor of $\sqrt{nw}$:
\ben \label{e19}
\wc{ds}_{string}^2 &=& - {f_1(r) \over f_3(r)} \, d\tau^2 + 
{f_2(r)\over f_3(r)} \, d\vec{y}^2
\, , \qquad r^2 = \vec y^2\, ,
\nonumber \\
\wc S &=& \sqrt{nw} \, f_3(r)\, , \nonumber \\
\wc T &=&  f_4(r)\, , \nonumber \\
\wc F^{(1)}_{r\tau} &=& f_5(r) \, , \nonumber
\\
\wc F^{(2)}_{r \tau} &=& f_6(r) \, .
\een
Using the inverse of the transformation \refb{e12} 
we can now generate the
modified version of the solution \refb{e10}:
\ben \label{e2.23a}
{ds}_{string}^2 &=& - {f_1(r) \over f_3(r)} \, d\tau^2 + 
{f_2(r)\over f_3(r)} \, d\vec{y}^2
\, , \qquad r^2 = \vec y^2\, ,
\nonumber \\
S &=& \sqrt{nw} \, f_3(r)\, , \nonumber \\
T &=&  \sqrt{n\over w} \, f_4(r)\, , \nonumber \\
F^{(1)}_{r\tau} &=& \sqrt{w\over n} \, f_5(r) \, , \nonumber
\\
F^{(2)}_{r \tau} &=& \sqrt{n\over w} \, f_6(r) \, .
\een

We now turn to the computation of entropy associated with this 
solution. In the presence of higher derivative corrections the entropy is 
no longer proportional to the area of the event horizon; there are 
additional corrections\cite{9307038,9403028,9502009}. These 
corrections all have the property that if the action is multiplied by a 
constant then the entropy associated with a given solution also gets 
multiplied by the same 
constant. Now suppose $a$ denote the entropy associated with the solution 
\refb{e17}. Then since the solution \refb{e17} and the action \refb{e15} 
are both universal, $a$ must be a purely numerical coefficient. Since 
\refb{e19} differs from \refb{e17} in a multiplicative factor of 
$\sqrt{nw}$ in the expression for $S$, and since from \refb{e15} we see 
that the effect of this multiplicative factor is to multiply the action by 
$\sqrt{nw}$, the entropy associated with the solution \refb{e19} must be 
given by\cite{9504147}:
\be \label{e20}
S_{BH} = a \sqrt{nw}\, .
\ee
Since \refb{e19} and \refb{e2.23a} are related by the transformation
\refb{e12} which is an exact symmetry of the tree level effective
action, \refb{e20} also gives the entropy associated with the
solution \refb{e2.23a}.

On the other hand counting of states of fundamental heterotic string 
carrying $w$ units of winding and $n$ units of momentum along $S^1$ shows 
that for large $n$ and $w$ the degeneracy of states grows as $e^{4\pi 
\sqrt{nw}}$. Thus the statistical entropy, defined as the logarithm of the 
degeneracy of states, is given by:
 \be \label{e21}
S_{stat} \simeq 4\pi\sqrt{nw} \, ,
\ee
for large $n$ and $w$. Thus we see that up to an overall 
multiplicative constant the 
statistical entropy agrees with the Bekenstein-Hawking entropy of the 
black hole\cite{9504147}.

For later use, it will be convenient to use
\refb{e9} to rewrite \refb{e2.23a} in terms of the original variables
$\rho$ and $t$:
\ben \label{e21a}
ds_{string}^2 &=& -{1\over g^2 nw}\, {f_1(g\rho)\over f_3(g\rho)}dt^2
+ g^2 \, {f_2(g\rho)\over f_3(g\rho)} \, d\vec x^2\, , 
\quad \rho=\sqrt{\vec x^2} \nonumber \\
S &=& \sqrt{nw} f_3(g\rho)\, , \nonumber \\
T &=& \sqrt{n\over w} \, f_4(g\rho)\, , \nonumber \\
F^{(1)}_{\rho t} &=& {1\over n} \, f_5(g\rho)\, , \nonumber \\
F^{(2)}_{\rho t} &=& {1\over w} \, f_6(g\rho)\, , \nonumber \\
ds_c^2 &=& S \, ds_{string}^2 = 
-{1\over g^2 \sqrt{nw}}\, f_1(g\rho)\, dt^2
+ g^2 \, \sqrt{nw} \, f_2(g\rho)\, d\vec x^2\,  .
\een

This finishes our review of \cite{9504147}. Let us now briefly 
mention the 
recent results of refs.\cite{0409148,0410076}. In 
these papers the authors compute the value of the coefficient $a$ by 
taking into account a special class of higher 
derivative terms in the effective action which are required for the 
supersymmetric completion of the curvature squared 
term\cite{rzwiebach} that is 
known to be 
present in the tree level effective action of the heterotic string 
theory\cite{9610237}. Based on earlier 
work\cite{9812082,9904005,9906094,9910179,0009234,9711053,0405146}
these papers 
concluded that in the presence of the higher derivative terms the 
solution near $r=0$ gets modified in such a way that the horizon acquires
a finite area. The naive
entropy computed from this using the Bekenstein-Hawking formula
is $2\pi\sqrt{nw}$.
However, as
was shown in \cite{0409148,0410076}, there are corrections to the entropy
formula due to the presence of the higher derivative terms in the action, and
these
give an additonal contribution of $2\pi\sqrt{nw}$. Thus 
the net entropy of the extremal black hole is given by $4\pi\sqrt{nw}$ in 
agreement with the statistical entropy \refb{e21}.

The analysis of \cite{0409148,0410076} was based on the assumption that
at the horizon the black hole solution develops enhanced supersymmetry.
While this leads to the solution close to the horizon, this does not
give us any information about the interpolating functions $f_1(r)$, $\ldots$
$f_6(r)$ for finite values of $r$.
In the next section we shall study the complete solution in the presence
of this special class of higher derivative terms, and find the functions
$f_1(r),\ldots f_6(r)$ which interpolate 
between the large $r$ limit discussed in this
section and the small $r$ results of refs.\cite{0409148,0410076}.

\sectiono{Modification of the Solution by Higher Derivative Terms and
its Near Horizon Limit} \label{scomplete}

In this section we shall find the modification of the solution
\refb{e6} by taking into account a special class of higher derivative
corrections to the effective action. In order to do so, we need to
first rewrite the low energy effective action \refb{e4} in the language
of $N=2$ supergravity and then analyze the effect of higher derivative
corrections.

\subsection{The low energy effective action as $N=2$ supergravity}
\label{s3.1}
The action of $N=2$ supergravity coupled to $n$
vector multiplets is governed by a prepotential $F$ which is a function
of $(n+1)$ complex scalars $X^I$ ($0\le I\le n$). The $X^I$'s are
projective coordinates and $F$ is a homogeneous function of the $X^I$'s
of degree 2. 
The gauge invariant bosonic degrees of freedom are the metric $g_{\mu\nu}$,
the complex scalars $X^I/X^0$, and a set of $(n+1)$ gauge fields
$\AAA^I_\mu$. Let us define
\be \label{e4.3.5}
\FF^I_{\mu\nu}\equiv \p_\mu \AAA^I_\nu - \p_\nu \AAA^I_\mu\, ,
\ee
\be \label{e4.4.5}
F_I \equiv {\p F \over \p X^I}\, , \qquad F_{IJ} \equiv {\p^2 F\over 
\p X^I \p X^J}\, ,
\ee
\be \label{e4.9}
L_I \equiv (F_{IJ} \bar X^J 
- \bar F_I), \qquad L \equiv 
\bar F - {1\over 2} \, F_{IJ} \, \bar X^I \bar X^J
\, , \qquad \NN_{IJ} \equiv {1\over 4} i \left( F_{IJ} +{1\over 2} 
{L_I L_J\over L}\right) 
\, ,
\ee
\be \label{e4.3}
e^{-\KK} \equiv i (\bar X^I F_I - X^I \bar F_I) \, ,
\ee
\be \label{e4.9a}
\FF^{I\pm}_{\mu\nu} \equiv {1\over 2} \, \left( \FF^I_{\mu\nu} \pm i 
\wt \FF^I_{\mu\nu}\right) \, .
\ee
\be \label{e4.9b}
\GG^{I\pm}_{\mu\nu} \equiv \pm \, 16 \pi \, i \, 
\left(\sqrt{-\det g}\right)^{-1}
 {\delta S\over 
\delta \FF^{I\pm}_{\mu\nu}}\, , 
\qquad \GG^I_{\mu\nu} 
\equiv \GG^{I+}_{\mu\nu} +
\GG^{I-}_{\mu\nu} \, ,
\ee
where $\wt \FF^I_{\mu\nu}$ denotes the Hodge dual of $\FF^I_{\mu\nu}$,
and in computing ${\delta S\over 
\delta \FF^{I\pm}_{\mu\nu}}$ we need to treat $\FF^{I\pm}_{\mu\nu}$
as independent variables.
Since $X^I$'s are projective coordinates,
we can impose a gauge condition
on the $X^I$'s. The convenient gauge choice is the
$e^{-\KK}=\hbox{constant}$ gauge. In this gauge 
the bosonic part of the
action takes the form\cite{oldsugra,9812082,0009234}:\footnote{In writing 
down this action we have implicitly assumed a reality condition on the 
fields such that the fields $X^I$ are either purely real or
purely imaginary and the prepotential $F$ is purely imaginary. In 
the present example this amounts to restricting the fields $S$ and $T$ 
defined in \refb{epre3} to be real. Otherwise there will be 
additional 
contribution to the kinetic term for the scalar fields. I would like to 
thank S.~Das for drawing my attention to this issue.}
\ben \label{e4.8}
\SSS &=& {1\over 8\pi} \int d^4 x \sqrt{-\det g} \bigg[ {1\over 2}
\, e^{-\KK} R 
- i g^{\mu\nu} (\p_\mu X^I \p_\nu \bar F_I - \p_\mu \bar X^I \p_\nu F_I)
\nonumber \\
&& + \left\{\NN_{IJ} g^{\mu\mu'} g^{\nu\nu'} \FF^{I-}_{\mu\nu} 
\FF^{J-}_{\mu'\nu'} + h.c. \right\} \bigg]\, ,
\een

For the system we are considering, the prepotential 
is\cite{9603191}
\be \label{epre1}
F = - {X^1 (X^2)^2\over X^0} \, .
\ee
If we define
the gauge invariant fields $S$ and $T$ through
\be \label{epre3}
{X^1\over X^0} = i \, S\, , \qquad {X^2\over X^0} = i \, T\, ,
\ee
and choose the gauge condition 
\be \label{egau1}
e^{-\KK}={1\over 2}\, ,
\ee
then from eqs.\refb{e4.3}, \refb{epre1} we get
\be \label{epre2}
X^0 = {1\over 4\, T\, \sqrt{S}}\, .
\ee
For real $S$ and $T$
eqs.\refb{e4.3.5} - \refb{e4.8} now produce
the action:
\ben \label{e4.10}
\SSS &=& {1\over 32\pi} \int d^4 x \sqrt{-\det g} \bigg[ R 
- {1\over 
2 
S^2} \, g^{\mu\nu} \, \p_\mu S \p_\nu S -  {1\over
T^2} \, g^{\mu\nu} \, \p_\mu T \p_\nu T 
\nonumber \\
&& - S T^2 \, 
g^{\mu\nu} \, g^{\mu'\nu'} \, \FF^{0}_{\mu\mu'} 
\FF^0_{\nu\nu'} - S^{-1}T^{2} \,
g^{\mu\nu} \, g^{\mu'\nu'} \, \FF^{1}_{\mu\mu'}
\FF^1_{\nu\nu'}  - 2 \, S \, g^{\mu\nu} \, g^{\mu'\nu'}
\, \FF^2_{\mu\mu'} \, \FF^2_{\nu\nu'}\, 
\bigg] \, . \nonumber \\
\een
This agrees with the action \refb{e4}
after a duality transformation on the field $\FF^1_{\mu\nu}$ 
if we make
the identification:\footnote{Note that heterotic string theory
compactified on $T^6$ has $N=4$ supersymmetry, but here we are
considering a truncated version of the theory which has $N=2$
supersymmetry.}
\be \label{e4.5}
F^{(1)}_{\mu\nu}= \FF^0_{\mu\nu} , \qquad F^{(2)}_{\mu\nu} 
 = \GG^1_{\mu\nu} = {T^2 \over S} \, \wt \FF^1_{\mu\nu}\, ,
\ee
and set $\AAA^2_\mu$ to 0. For a general configuration of
$X^I$'s and $\AAA^I_\mu$'s the 
imaginary parts of $T$ and $S$ can be identified 
respectively with
an appropriate off-diagonal component of the internal metric
and the axion field obtained by dualizing the field $B_{\mu\nu}$,
whereas $\AAA^{2}_{\mu}$ can be regarded
as an appropriate linear combination 
of $G^{(10)}_{m\mu}$ and $B^{(10)}_{m\mu}$ for $5\le m \le 9$.
During the rest of our analysis we shall consider
configurations where $S$ and $T$ are real.

\renewcommand{\wh}{\widehat}

\subsection{Higher derivative corrections} \label{s3.2}
The corrections associated with supersymmetrization 
of the curvature squared 
term can be taken into account by modifying the
prepotential to\cite{9812082,0009234}
\be \label{e4.1}
F = - {X^1 (X^2)^2\over X^0} + \, \wh A \, f\left({X^1\over X^0}
\right)\, ,
\ee
where $\wh A$ is a background chiral superfield whose highest
component contains
the square of the Weyl tensor, and $f$ is a function to be
specified later (see eqs.\refb{e5.7} 
and \refb{e7.5} below). It has the property
\be \label{efreal}
f(iS) + (f(iS))^*= 0 \quad \, \hbox{for real $S$}\, .
\ee
We define
\be \label{e4.4}
F_{\wh A} = {\p F \over \p \wh A}=f\left({X^1\over X^0}\right)\, ,
\ee
and various other quantities as in eqs.\refb{e4.3.5}-\refb{e4.9b}.
An expression for the bosonic part of the action for a general 
prepotential
$F(X^I, \wh A)$ has been given in \cite{0009234}, but we shall
not review it here. 
Instead we shall focus our attention on a class
of $N=1$ supersymmetric black hole solutions constructed in
\cite{0009234,0012232} after taking into account the 
corrections
given in \refb{e4.4}.
In an appropriate gauge these solutions have the 
form:\footnote{Ref.\cite{0009234} considered a more general class
of solutions by allowing the supersymmetry transformation 
parameter to rotate by a phase as we move in space. Since we shall
be interested in a solution for which the fields $S$ and $T$ are real
we shall set the phase to 1.}
\be \label{e4.11}
ds_c^2 = -e^{2G(\rho)} dt^2 + e^{-2G(\rho)} d\vec x^2\, , 
\qquad \rho=\sqrt{\vec x^2}\, ,
\ee
\be \label{e4.12}
e^{-G} (X^I - \bar X^I) = i\left(a_I + {b_I\over \rho}\right)\, ,
\ee
\be \label{e4.13}
e^{-G}(F_I - \bar F_I) = i\left(c_I +{d_I\over \rho}\right)\, ,
\ee
where $a_I$, $b_I$, $c_I$, $d_I$ are arbitrary real 
constants,\footnote{Physically the constants $a_I$ and $c_I$ measure
the asymptotic values of various fields whereas $b_I$ and $d_I$ measure
the charges carried by the black hole.}
\be \label{e4.13a}
\FF^I_{\rho t} = \p_\rho (e^G(X^I+\bar X^I))\, ,
\ee
\be \label{e4.13b}
\GG^I_{\rho t} = \p_\rho (e^G (F_I+\bar F_I))\, ,
\ee
\be \label{e4.14}
\wh A = -64 \, e^{2G} \, (\p_\rho G)^2\, ,
\ee
\be \label{e4.15}
e^{-\KK} +{1\over 2}
\chi = - 128 i e^{3G} {1\over \rho^2} \p_\rho \left(\rho^2 e^{-G} 
\p_\rho G \,
(F_{\wh A} - \bar F_{\wh A})\right)\, .
\ee
Here $\chi$ is an arbitrary constant whose value is determined
by the gauge condition. We shall choose the gauge
\be \label{e4.16}
\chi = -1\, ,
\ee
so that in the absence of coupling to the background superfield $\wh A$
the gauge condition agrees with \refb{egau1}.

The procedure for solving these equations is as follows. For given 
constants $a_I$, $b_I$, $c_I$, $d_I$, eqs. \refb{e4.12}, \refb{e4.13}
give $2n$ real equations which can be used to used to solve for
the $n$ complex $X^I$'s in terms of $G$ and $\wh A$. \refb{e4.14}
gives $\wh A$ in terms of $G$. Substituting these in \refb{e4.15}
we get a differential equation for $G$ which can then be solved.
Once $G$ and the $X^I$'s have been found, we can use \refb{e4.13a},
\refb{e4.13b}
to calculate the gauge field strengths $\FF^I_{\rho t}$ and 
$\GG^I_{\rho t}$.

Thus in order to find the black hole solution describing the elementary
string states, we first need to determine the constants $a_I$, $b_I$,
$c_I$ and $d_I$. For this we note that in the absence of the coupling
to the background superfield $\wh A$, \i.e. for 
$f(X^1/X^0)=0$, the solution
\refb{e6} has the form given
in eqs.\refb{e4.11}-\refb{e4.16} for the 
following choice
of the constants:
\ben \label{e5.3}
&& a_1 = 2g^{-1} R^{-1}, \quad b_1 = w, \quad c_0 = -{R\over 8} g^{-1},
\quad d_0 = -n, \quad a_2 ={1\over 2} \, g\, , \nonumber \\
&& a_0=b_0=b_2=0, \qquad c_I=d_I=0 \quad
\hbox{for} \quad I=1,2\, .
\een
In order to study the modification of the solution due to coupling to
the background superfield $\wh A$, we first note that for a given 
solution the constants $a_I$, $b_I$, $c_I$, $d_I$ 
may be determined by
knowing the form of the solution at large $\rho$ to
order $1/\rho$. As argued in the last
section, the modification of the solution due to the four and higher
derivative terms in the action appear at order $1/\rho^2$. Hence
the constants $a_I$, $b_I$, $c_I$ and $d_I$ should not
change due to the higher derivative corrections and must have
the same values as given in
eq.\refb{e5.3}. With this choice $X^0$ is real
and $X^1$, $X^2$ are purely imaginary, and the
non-trivial components of 
eqs.\refb{e4.11}-\refb{e4.14} may be expressed as:\footnote{We continue 
to identify $S$, $T$, $F^{(1)}_{\mu\nu}$ and $F^{(2)}_{\mu\nu}$ with
$-iX^1/X^0$, $-iX^2/X^0$, $\FF^0_{\mu\nu}$ and $\GG^1_{\mu\nu}$
respectively, but the fields defined this way may no longer be related
to the ten dimensional fields via eqs.\refb{e3}, \refb{e4a}.
We shall elaborate
on this in section \ref{s3.4}.}
\ben \label{e5.4}
e^{-G} \, X^0 \, S &=& {1\over 2} \, \left( 2 g^{-1} R^{-1} + 
{w\over \rho}\right) \nonumber \\
e^{-G} \, X^0 \, T &=& {g\over 4} \, , \nonumber \\
e^{-G} \, X^0 \, S \, \left( T^2 + {\wh A\over (X^0)^2} \, 
f'(iS)\right) &=& {1\over 2} \left( {R\over 8} g^{-1} 
+{n\over \rho}\right) \, ,
\een
\be \label{e5.4a}
\wh A = -64 \, e^{2G}\, (\p_\rho G)^2\, ,
\ee
\ben \label{e5.55}
F^{(1)}_{\rho t} &=& \FF^0_{\rho t} = 2\p_\rho\left(e^G X^0\right),
\nonumber \\
F^{(2)}_{\rho t}
&=& \GG^1_{\rho t} = 2\p_\rho\left[e^G X^0\left( T^2 + {\wh 
A\over (X^0)^2} \, f'(iS)\right)\right]\, .
\een
For the choice of prepotential given in \refb{e4.1}, we have,
using \refb{efreal},
\be \label{e5.5}
e^{-\KK} \equiv i(\bar X^I F_I - X^I \bar F_I)
= 4 \, S\, \left(2 \, T^2 + \, {\wh A\over (X^0)^2} 
f'(iS)\right) \, (X^0)^2 \, .
\ee
Eqs.\refb{e4.15}, \refb{e4.16} now give:
\be \label{e5.6}
4 \, S\, \left(2 \, T^2 + \, {\wh A\over (X^0)^2} 
f'(iS)\right) \, (X^0)^2 = {1\over 2} - 256\, i \, e^{3 G} \, 
{1\over \rho^2} \, \p_\rho \left[ \rho^2 \, e^{-G}\, \p_\rho G\,
f(iS)\right]\, .
\ee
In order to solve these equations, we can first use eqs.\refb{e5.4},
\refb{e5.4a} to express $S$, $T$, $X^0$
and $\wh A$ in terms of $G$, and
then substitute these into \refb{e5.6} to get a 
second order non-linear differential 
equation for $G$.

We shall now study various aspects of these equations.

\subsection{Near horizon geometry and entropy} \label{s3.3}
In the $\rho\to 0$ limit we can 
rewrite the first two equations in \refb{e5.4} as:
\ben \label{e5.0a}
X^0 &\simeq& {w\over 2\rho} \, e^G \, S^{-1}\, , \nonumber \\
T &\simeq& {g\over 2w} \, \rho \, S\, .
\een
Using \refb{e5.4a} and \refb{e5.0a} the last equation of 
\refb{e5.4} now gives
\be \label{e5.0b}
{\rho^2 \over w^2} \, S^2 \left[{g^2\over 4} - 256 \, 
(\p_\rho G)^2 \, f'(iS)\right] \simeq {n\over w}\, .
\ee
On the other hand eq.\refb{e5.6} takes the form:
\be \label{e5.0c}
S \, e^{2G} \, \left[ {g^2\over 2} - 256 \, (\p_\rho G)^2 
f'(iS)\right]
= {1\over 2} - 256 \, i\, e^{3G} \, {1\over \rho^2} \, \p_\rho\left[
\rho^2 \, e^{-G} \, \p_\rho G \, f(iS)\right] \, .
\ee
If we take the following ansatz for the solutions near $\rho=0$
\be \label{e5.0d}
S \simeq S_0, \qquad e^{2G} \simeq K_0 \rho^2\, ,
\ee
then by substituting this into \refb{e5.0b}, \refb{e5.0c} we get
\be \label{e5.0e}
S_0^2 \, f'(iS_0) = -{1\over 256} \, nw\, , \qquad S_0 \, K_0 
\, f'(iS_0) = -{1\over 512}\, .
\ee
For a given function $f$ these equations can be solved to find $S_0$
and $K_0$.
Using \refb{e5.0a}, \refb{e5.4a}, \refb{e5.55} and 
\refb{e5.0e} we get
\be \label{e5.0f}
X^0\simeq {w\over 2} \, K_0^{1/2} \, S_0^{-1}\, , \qquad T\simeq
{g\over 2 w} \, S_0 \, \rho  \, , \qquad F^{(1)}_{\rho t} \simeq 
{1\over 2n}\, , \qquad 
F^{(2)}_{\rho t}\simeq {1\over 2w}\, .
\ee
This determines the field configuration near $\rho=0$.

Substituting \refb{e5.0d} into the expression for the metric
given in \refb{e4.11} we see that the 
area of the event horizon, measured in the canonical metric, is
\be \label{e5.0g}
A_H = 4\, \pi \, K_0^{-1} \, .
\ee
Thus the naive black hole entropy will be given by 
\be \label{e5.0i}
{A_H\over 4G_N}={\pi\over 2} K_0^{-1}\, ,
\ee 
where we have used $G_N=2$ as given in \refb{e5.0h}.
However as shown in \cite{9812082}, due
to the presence of higher derivative terms in the action 
this expression gets modified to
\be \label{e5.0j}
S_{BH} = {A_H\over 4 G_N} - 256 \, \pi \, Im(F_{\wh A})\, ,
\ee
where $F_{\wh A}$ has been defined in \refb{e4.4}. Using the expression
\refb{e4.1} for $F$, and eqs.\refb{e5.0g}, \refb{e5.0h}, \refb{e5.0e}
we can express
\refb{e5.0j} as:
\be \label{e5.0k}
S_{BH} = {1\over 2} \, \pi \, K_0^{-1} \, - \, 256 \, \pi \,
Im(f(iS_0)) = -256 \, \pi \, \left( S_0 f'(iS_0) + 
Im (f(iS_0))\right)\, .
\ee
Thus once $K_0$ and $S_0$ have been determined from \refb{e5.0e},
eq.\refb{e5.0k} can be used to compute the 
black hole entropy. Note that
although eqs.\refb{e5.4}-\refb{e5.6} 
represent the condition for preserving
half of the space-time supersymmetries 
of the vacuum, the solution
\refb{e5.0e}, \refb{e5.0f} at the horizon $\rho=0$
actually preserves larger 
number of supersymmetries\cite{9812082,0009234}.

There is however a subtle point that we have overlooked. The
remark below \refb{e5.6} shows that $G$ satisfies a second order
non-linear differential equation. \refb{e5.0d}, \refb{e5.0e}
describes a particular solution of this equation near $\rho=0$.
In order to show that this describes the correct behaviour of the
{\it black hole solution} near $\rho=0$, we need to ensure that
this solution approaches the correct asymptotic form 
\refb{e6} for
large $\rho$ where the effect of higher derivative corrections
should be negligible. Since a general solution of the differential
equation has two integration constants, there is no {\it a priori}
guarantee that the choice of integration constants which lead
to the form given in \refb{e5.0d}, \refb{e5.0e} will also have the
correct asymptotic behaviour. We shall return to this issue in
section \ref{s3.5a}.

\subsection{T-duality} \label{s3.4}
The near horizon expression for $T$ 
given in \refb{e5.0f} shows 
that it vanishes as $\rho\to 0$. If $G^{(10)}_{44}$ 
is identified with $T^2$ as in eq.\refb{e3}, 
then this would imply that 
$G^{(10)}_{44}$ would vanish as $\rho\to 0$. This is somewhat surprising 
if we consider the fact that before including the higher derivative 
corrections to the action, $G^{(10)}_{44}$ approached a finite value 
$T^2={n\over w}$ as 
$\rho\to 0$ (see eq.\refb{e8}). 
Since unlike the field $S$, $G^{(10)}_{44}$ 
had already reached a 
fixed point 
value, one would have expected that higher derivative corrections would 
not drastically modify this behaviour.

We shall now argue that after inclusion of the higher derivative 
corrections given in \refb{e4.1}, the correct identification of 
$G^{(10)}_{44}$ is not $T^2$, but,
\be \label{e6.1}
G^{(10)}_{44} = T^2 + {\wh A\over (X^0)^2} \, f'(iS)\, .
\ee
In that case the first and the last equations in
\refb{e5.4} show that in the $\rho\to 0$ limit
\be \label{e6.2}
G^{(10)}_{44} \to {n\over w}\, .
\ee
This agrees with the near horizon value of $G^{(10)}_{44}$ in the 
absence of higher derivative corrections.

In order to establish \refb{e6.1} we need to study how the T-duality
transformation 
\be \label{e6.2a}
G^{(10)}_{44} \to (G^{(10)}_{44})^{-1}\, ,
\ee
which is an
exact symmetry of heterotic string theory on $T^5\times S^1$, 
is realized in terms of
the fields $X^I$. According to \cite{9603191} 
this corresponds to the
transformation:
\ben \label{e6.3}
X^0 &\to& \wt X^0 = - F_1 = {(X^2)^2\over X^0} - {\wh A\over X^0} 
\, f'(iS) \, , \nonumber \\
X^1 &\to& \wt X^1 = F_0 = {X^1\, (X^2)^2\over (X^0)^2} - 
{\wh A\, X^1\over (X^0)^2} 
\, f'(iS) \, , \nonumber \\
X^2 &\to& \wt X^2 = -X^2   \, ,
\een
\ben \label{e6.3a}
F_0 &\to& \wt F_0 = X^1 \, , \nonumber \\
F_1 &\to& \wt F_1 = -X^0 \, , \nonumber \\
F_2 &\to& \wt F_2 = - F_2 = -{2 X^1 X^2\over X^0}\, ,
\een
where
\be \label{e6.4}
\wt F_I \equiv F_I(\{\wt X^I\}, \wh A)\, .
\ee
Eqs.\refb{e6.3} describes the transformation laws of the fields $X^I$,
whereas eqs.\refb{e6.3a} are consistency conditions which must be
satisfied in order that the transformations \refb{e6.3} are symmetries
of the equations of motion. It can be easily verified that 
eqs.\refb{e6.3a} follow from eqs.\refb{e6.3}.

Using \refb{e6.3} we see that
\ben \label{e6.5}
&& \wt S = -i\, {\wt X^1\over \wt X^0} = -i\, {X^1\over X^0} = S\, , 
\nonumber \\
&& \wt T = -i {\wt X^2 \over \wt X^0} = i {X^2 \over
{(X^2)^2\over X^0} - {\wh A\over X^0} 
\, f'(iS) } = {T \over T^2 + {\wh A\over (X^0)^2} \, f'(iS)}\, .
\een
In the absence of higher derivative corrections, \i.e. when 
$f(iS)=0$, this gives the familiar $T\to T^{-1}$ duality
transformation. However we see that the duality transformation
law of $T$ gets modified by the higher derivative terms. Thus
$T$ can no longer by identified as $\sqrt{G^{(10)}_{44}}$ which
transforms as \refb{e6.2a} even when higher derivative corrections
are included. On the other hand we note from \refb{e6.3} that
\be \label{e6.6}
\wt T^2 + {\wh A\over (\wt X^0)^2} \, f'(i\wt S) = {1 \over
T^2 + {\wh A\over (X^0)^2} \, f'(iS)}\, .
\ee
Comparing this to \refb{e6.2a}, and by using the requirement that for
$f(iS)=0$, $G^{(10)}_{44}$ should reduce to $T^2$, 
we reach the identification given in
\refb{e6.1}.
 
\subsection{Tree level heterotic string theory and universality}
\label{s3.5}
The higher 
derivative corrections to the tree level 
effective action
of heterotic string theory are given by the following choice of the
function $f(u)$\cite{0409148}:
\be \label{e5.7}
f(u) = -{C\over 64} \, u\, , \qquad C=1\, .
\ee
Although the constant $C$ is equal to unity, we shall analyze the solution
assuming that it is an arbitrary constant so that at various stages we can
recover the leading $\alpha'$ result by setting $C$ to 0.
Eq.\refb{e5.0e} now gives:
\be \label{e5.7a}
S_0={1\over 2}\, \sqrt{nw\over C}\, , \qquad K_0 
= {1\over 4\sqrt{C\, nw}}\, .
\ee
Using eqs.\refb{e5.0k} and \refb{e5.7a} we get:
\be \label{e5.7b}
S_{BH} = 4\, \pi\, \sqrt{C\, nw}\, .
\ee
For $C=1$ 
this reproduces the microscoic entropy \refb{e21} in agreement with 
the results of \cite{0409148}. Setting $C=0$ we recover the supergravity
result that the entropy vanishes.

We shall now explicitly check that this solution reproduces the
scaling property encoded in eq.\refb{e21a} in the limit of large $n$
and $w$. For this we take the limit of large $n$ and $w$ in
eqs.\refb{e5.4} - \refb{e5.6}, substitute the general form \refb{e21a}
into these equations, and use the form 
\refb{e5.7} for $f(u)$.
First of all comparison between the form of the metric \refb{e21a}
and \refb{e4.11} gives
\be \label{e5.8}
f_1(r) f_2(r) = 1\, .
\ee
Eqs.\refb{e5.4}-\refb{e5.6} in this limit give
\be \label{e5.9}
X^0 = {1\over 2g\rho} \, \sqrt{w\over n} \, (nw)^{-1/4} \, 
{\sqrt{f_1(g\rho)}\over f_3(g\rho)}\, ,
\ee
\ben \label{e5.10}
&& {f_3(r)\over f_4(r)} = {2\over r} \, , \nonumber \\
&& f_3(r) \, f_4(r) \, \left[ 1 + 4\, C\, \left( {f_1'(r)
\over f_1(r)}\right)^2\right] 
= {2\over r} \, , \nonumber \\
&&{1\over 2} \, f_1(r) \, f_3(r) \, = {1\over 2} - {2\, C\over r^2}
f_1(r) \, \p_r 
\, \left[ r^2 \, f_3(r) \, {f_1'(r)\over f_1(r)}\right]\, ,
\nonumber \\
&& f_5(r) = \p_r \left({1\over r} \, {f_1(r) \over f_3(r)}\right)\, ,
\nonumber \\
&& f_6(r) = \p_r \left({1\over r} \, {f_1(r) \over f_3(r)}\right)\, .
\een
We see that eqs.\refb{e5.8} and \refb{e5.10} involving
the functions $f_i(r)$ are completely independent of any external
parameters, as predicted by the scaling argument.

We can try to solve these equations by introducing
a new function $h(r)$
through
\be \label{e7.1}
f_1(r) = e^{h(r)}\, .
\ee
Then \refb{e5.8} and the first two and the last two 
equations of \refb{e5.10} give\footnote{We could also have 
gotten eqs.\refb{e7.1}-\refb{e7.3} by directly substituting the
general form \refb{e17} into eqs.\refb{e4.11}-\refb{e4.16} 
with $(\rho,\vec x,
t)$ replaced by $(r, \vec y, \tau)$, and $a_I$, $b_I$, $c_I$, $d_I$
determined from the asymptotic form \refb{e18}. In this case the
$n$, $w$, $g$ and $R$ independence of the resulting equations
would be manifest from the beginning.}
\ben \label{e7.2}
f_2(r) &=& e^{-h(r)} \, , \nonumber \\
f_3(r) &=& {2\over r} \, {1\over \sqrt{1 + 4 \, C\, (h'(r))^2}}
\, , \nonumber \\
f_4(r) &=&  {1\over \sqrt{1 + 4\, C\, (h'(r))^2}}
 \, , \nonumber \\
f_5(r)&=& {1\over 2} \, \p_r \left( e^{h(r)} \, 
\sqrt{1 + 4 \, C\, (h'(r))^2} \right) \, , \nonumber \\
f_6(r) &=& {1\over 2} \, \p_r \left( e^{h(r)} \, 
\sqrt{1 + 4 \, C \,(h'(r))^2} \right) \, . \nonumber \\
\een
Finally, substituting these into the third equation of
\refb{e5.10} we get a differential equation for $h$:
\be \label{e7.3}
C\, h' \left(1 + 4\, C\, (h')^2\right) + C \, r \, h'' 
= {r^2\over 8} \, e^{-h} \, \left(1 + 4\, C\, (h')^2\right)^{3/2}
- {r\over 4} \, \left(1 + 4\, C\, (h')^2\right) \, .
\ee
The boundary condition on $h$ follows from \refb{e18}
\be \label{e7.4}
e^h \simeq {r\over 2} \, \quad \hbox{for large $r$}\, .
\ee
One can easily verify that \refb{e7.4} satisfies \refb{e7.3}
for large $r$.

For small $r$ eq.\refb{e7.3} admits a solution
\be \label{e8.0a}
e^h \simeq {r^2\over 4 \sqrt C}\, ,
\ee
which leads to the solution \refb{e5.7a}. 
However since
\refb{e7.3} is a second order differential equation for $h$, there
is no {\it a priori} guarantee that there is a smooth
solution that interpolates between \refb{e7.4} and \refb{e8.0a}.
We shall analyze this issue in section \ref{s3.5a}.
Note that for $C=0$ eq.\refb{e7.3} becomes a
purely algebraic equation for $h$ which admits a unique  solution
$e^h = r/2$ and reproduces the result of section \ref{sreview}.

\renewcommand{\wh}{\hat}

\subsection{The analysis of the interpolating solution} \label{s3.5a}

\begin{figure}
\leavevmode
\begin{center}
\epsfysize=2in
\epsfbox{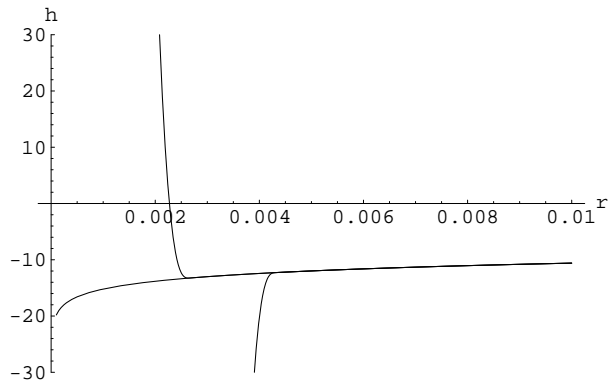}
\end{center}
\caption{Trajectories neighbouring $h=2 \ln{r\over 2}$ for small $r$.}
\label{ff1}
\end{figure}

We shall now analyze the differential equation 
\refb{e7.3} for $C=1$ and analyze
the possibility of a solution that interpolates between \refb{e7.4} for
large $r$ and \refb{e8.0a} for small $r$. We shall begin by analyzing
fluctuations around the asymptotic solutions. For large $r$,
if we make the ansatz
\be \label{e11.1}
h(r) = \ln {r\over 2} + \phi\, ,
\ee
and assume that $\phi$ and all its derivatives are of order unity,
then \refb{e7.3} gives, for $C=1$,
\be \label{e11.2}
\phi'' = {1\over 4}\, e^{-\phi} \, \{ 1 + 4(\phi')^2\}^{3/2}
-{1\over 4} \, \{ 1 + 4 (\phi')^2\}\, .
\ee
$\phi=0$ is a solution of this equation as expected. For small
$\phi$ the equation reduces to
\be \label{e11.3}
\phi''=-{1\over 4} \, \phi +\OO(\phi^2)\, ,
\ee
which has, as solutions
\be \label{e11.4}
\phi = A \, \cos\left({r\over 2}+ B\right) + \OO(A^2)\, ,
\ee
for arbitrary integration constants $A$, $B$. Thus $\phi=0$ is an elliptic
fixed point of the second order autonomous system described by \refb{e11.2}
and we expect that
there is a generic set of initial conditions for which the solution to
\refb{e7.3} at large $r$ will have periodic 
oscillations around $h=\ln{r\over 2}$:
\be \label{e11.5}
h = \ln{r\over 2} +A \, \cos\left({r\over 2}+ B\right) +\OO(A^2)\, .
\ee
 
Analyzing the fluctuations of the solutions around the
solution \refb{e8.0a} near $r=0$ is more difficult. Numerical analysis
suggests that the behaviour of the solution around $r=0$ is highly 
unstable and for slight changes in the initial condition the solution
develops spontaneous singularities at some value of $r$ close to zero.
This has been illustrated in Fig.\ref{ff1} where we have displayed some
trajectories neighbouring the solution \refb{e8.0a} for small $r$.

\begin{figure}
\leavevmode
\begin{center}
\epsfysize=2in
\epsfbox{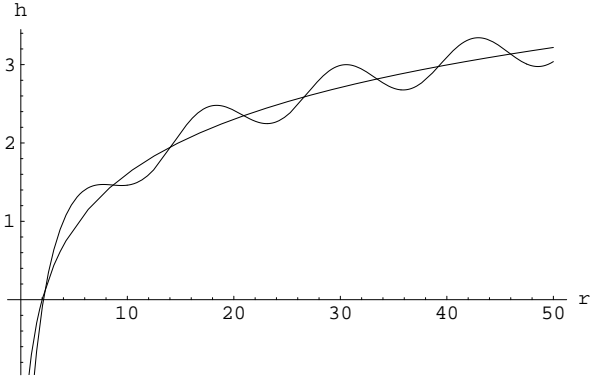}
\end{center}
\caption{Numerical result for the solution to \refb{e7.3}
satisfying the boundary
condition $h = 2\ln{r\over 2}$ for small $r$.
The smooth curve represents $h=\ln{r\over 2}$.}
\label{ff2}
\end{figure}

For this reason, in order to study if there is a solution to \refb{e7.3}
that interpolates between \refb{e7.4} and \refb{e8.0a}, we begin with the
solution \refb{e8.0a} for small $r$ and numerically integrate it to 
study its behaviour at large
$r$. The result is shown in Fig.\ref{ff2}. We see from this 
that the solution
does not approach \refb{e7.4}, but oscillates around it, 
as is expected for a generic initial condition. This seems unlikely to
be a numerical error, and seems to 
indicate that the solution that has the correct near 
horizon behaviour does not approach the desired form at large $r$. 

\renewcommand{\wh}{\hat}
We 
shall now argue however that there is a subtlety in this interpretation 
and that once this subtlety is taken into account, the asymptotic 
behaviour of the solution is consistent with the desired form.
For this we note that for small $A$, 
the solution \refb{e11.4} implies the following asymptotic forms for the 
$f_i$'s:
\ben \label{e12.1}
&& f_1 \simeq {r\over 2} \left( 1 + 
A \cos\left({r\over 2} + B\right) 
\right), \quad f_2 \simeq {2\over r} 
\left( 1 - A \cos\left({r\over 2} + 
B\right)\right), \quad f_3\simeq {2\over r}, 
\quad f_4 \simeq 1\, , \nonumber \\
&& f_5=f_6 \simeq {1\over 4}-{Ar\over 8} \sin\left({r\over 2} 
+ B\right) +{A\over 4}\, \cos\left({r\over 2}+B\right)\, .
\een
Substituting these into \refb{e17} we see that at large $r$, and to 
linear order in $A$, the 
modification of  the solution appears only in the expression for the 
metric and the gauge fields. 
In particular, we have
\ben \label{e12.2}
&& 
\wh{ds}_{string}^2 \simeq -{r^2 \over 4} \left( 1 + A \cos\left({r\over 2} 
+ 
B\right)\right) d\tau^2 + \left( 1 - A \cos\left({r\over 2} +
B\right)\right) \, d\vec x^2\, , \nonumber \\
&& \wh F^{(1)}_{rt} = \wh F^{(2)}_{rt} \simeq {1\over 4}-{Ar\over 8}\, 
\sin\left({r\over 2} + B\right)+{A\over 4}\, 
\cos\left({r\over 2}+B\right)\, .
\een
Under the change of variable $\tau = 2\eta/r$, the metric and the 
gauge fields take the form  
\ben \label{e12.3}
&& \wh{ds}_{string}^2 = -d\eta^2 + d\vec x^2 - A \cos\left({r\over 2} +
B\right) (d\eta^2 + d\vec x^2) +\OO\left({1\over r}\right)
\nonumber \\
&&\wh F^{(1)}_{r\eta} = \wh F^{(2)}_{r\eta} \simeq -{A\over 4}\, 
\sin\left({r\over 2} + B\right) +\OO\left({1\over r}\right)\, .
\een

This shows that for small $A$ the
asymptotic solution differs from the (locally)
flat background by 
an 
oscillatory piece proportional to $A$. Hence this must represent a 
solution of the linearized equations of motion.\footnote{Although
\refb{e7.2}, \refb{e7.3} 
were derived using the requirement of supersymmetry
preservation, it has been argued in \cite{0009234} that a solution of
these equations also satisfy the classical field equations.
We have checked explicitly that the metric fluctuations given in
\refb{e12.3} 
does satisfy the linearized equations of motion around the flat
background, but we shall not demonstrate it here.}
This might seem surprising 
since normally the only solution of linearized equations of motion for the 
graviton and the gauge fields
are gravitational and electromagnetic
wave solutions. However in the present 
circumstances there can be additional solutions because 
the action has higher 
derivative terms. In order to illustrate this we consider the 
simpler example of a 
scalar field $\psi$ with action:
\be \label{e12.4}
{1\over 2} \int d^4 x \, \psi \, \square \left(1 - {\square\over 
M^2}\right) \, \psi\, .
\ee
The equations of motion for $\psi$
has solutions of the form $A \, e^{i k.x}$ 
with
\be \label{e12.5}
k^2 = 0 \quad \hbox{or} \quad - M^2\, .
\ee
Thus a single scalar field can describe plane waves of different masses in 
the presence of higher derivative terms. Similar phenomenon occurs for 
gravity and gauge fields
in the presence of higher derivative terms in the action.

Note however that the presence of such additional oscillatory
solutions 
will, upon quantization, give rise to
additional quantum states 
which are not present in the spectrum of 
string theory. Thus there is an apparent contradiction between field 
theory and string theory results. This problem was resolved by 
Zwiebach\cite{rzwiebach} who argued that these higher derivative terms
should
be removed by appropriate field redefinition. For example by making a 
field redefinition 
\be \label{e12.6}
g_{\mu\nu} \to g_{\mu\nu} + a \, R_{\mu\nu} + b \, R \, g_{\mu\nu}\, ,
\ee
for appropriate constants $a$ and $b$, we can ensure that the curvature 
squared terms appear in the action in the Gauss-Bonnet combination. This 
particular combination of terms has the property that when we expand this 
in the weak field approximation, the quadratic term involving the graviton 
field does not receive any contribution from the curvature squared term. 
As a result the linearized equations of motion of the graviton field 
remain unmodified and we only get the usual plane wave solutions. Since 
this is what string spectrum predicts, we see that this redefined metric 
is the correct 
variable to be used to make direct contact with string theory. A similar
field redefinition 
must be carried out for the gauge fields as well. 

Under such
field redefinitions the oscillatory solutions of the type 
given in \refb{e12.3} are mapped to zero.
We shall illustrate this in the context of the scalar field action 
\refb{e12.4}. The field redefinition that brings the action to the 
standard action for a massless scalar field is
\be \label{e12.7}
\wt \psi =  \left(1 - {\square\over
M^2}\right)^{1/2} \psi\, .
\ee
Under this map the solution $\psi=Ae^{ik.x}$ gets mapped to $Ae^{ik.x}$ 
for $k^2=0$ and to 0 for $k^2=-M^2$. Thus in terms of the variable $\wt 
\psi$ only the plane wave solutions with $k^2=0$ are present.

This discussion shows that the fluctuations proportional to $A$ in 
\refb{e12.3} are unphysical and are in fact mapped to zero when we use the 
correct field variables. It should be emphasized that we have not
explicitly constructed the field redefinition, but are relying on the
fact that the effective field theory that correctly describes tree
level string theory must admit such field redefinitions.
Although our discussion has been focussed at the 
linearized level, we expect that the result should be valid beyond the 
linear approximation, and that when we use the right field variables, the 
two parameter family of solutions of the differential equation 
\refb{e11.2}, valid for large $r$, will map to a single solution. This in 
turn would imply that 
the oscillations that we see in Fig.\ref{ff2} are due to the wrong 
choice 
of field variables, and should disappear once we make the right choice. 
Presumably when we use the right choice of field variables the
differential equation \refb{e7.3} will be replaced by an ordinary
equation with a unique solution which will interpolate correctly
between the desired asymptotic limits.

\subsection{Effect of quantum corrections and holomorphic
anomaly} \label{s3.6}

Finally let us consider the contribution to the black hole 
entropy obtained after taking into account the full quantum
corrections to the function $f(u)$.\footnote{Note however that
this takes into account only a special class of corrections and
does not correspond to the full quantum corrected black hole 
entropy.}$^,$\footnote{Ref.\cite{0409148} follows a somewhat
different approach for relating the quantum corrected black hole
entropy to the statistical entropy. This uses a mixed 
ensemble\cite{0405146} and does not explicitly take into
account the effect of holomorphic anomaly.
Presumably the two approaches
are related but the relationship is not completely clear to us.}
In this case\cite{9610237}
\be \label{e7.5}
f(u) = -{1\over 128 \, \pi\, i} \ln \Delta
\left(e^{2\pi i u}\right)\, ,
\ee
where
\be \label{e7.6}
\Delta(q) = (\eta(q))^{24}, \qquad
\eta(q) = q^{1\over 24} \, \prod_{n=1}^\infty \, (1 - q^n) \, .
\ee
Eqs.\refb{e5.0e}, \refb{e5.0k} now give
\be \label{e9.1}
S_0^2 \, {\Delta'\left(e^{-2\pi S_0}\right) \over
\Delta\left(e^{-2\pi S_0}\right)} = {nw\over 4}\, , \qquad
S_{BH} = 4\pi \left[ S_0 \, {\Delta'\left(e^{-2\pi S_0}\right) \over
\Delta\left(e^{-2\pi S_0}\right)} \, - \, {1\over 2\pi}
\, \ln \Delta\left(e^{-2\pi S_0}\right) \right]\, ,
\ee
where
\be \label{ethirda}
\Delta'(q)\equiv q {\p \Delta(q)\over \p q}\, .
\ee
Note that for small $q$
\be \label{e7.7}
\ln (\Delta(q)) = \ln q + \OO(q)\, .
\ee
This gives, for large $S$,
\be \label{e7.8}
f(iS) \simeq -{i\over 64} \, S + \OO\left(e^{-2\pi S}\right)\, .
\ee
This agrees with \refb{e5.7} for large $S$. Thus for large $S_0$ the
quantum corrected answer for the entropy, computed 
from eqs.\refb{e5.0e}, 
\refb{e5.0k}, reduces to the tree level answer \refb{e5.7a}, 
\refb{e5.7b}, as is
expected. We can take into account
the corrections by solving the equations for $K_0$
and $S_0$ iteratively as a power series expansion in $e^{-2\pi S_0}$.
Since to leading order $S_0$ is given by ${1\over 2}\sqrt{nw}$ we see
that the quantum corrections to the black hole entropy from this
special class of higher derivative terms is of order 
$e^{-\pi\sqrt{nw}}$ for large $nw$. Thus we have
\be \label{e7.8a}
S_{BH} = 4\, \pi\, \sqrt{nw} + \OO\left(e^{-\pi\sqrt{nw}}
\right)\, .
\ee

This however is not the complete story.\footnote{I would like
to thank R.~Gopakumar for drawing my attention to the role of
holomorphic anomaly in producing logarithmic corrections.}
As was pointed out in
\cite{9904005,9906094}, there are corrections to the above formula
due to holomorphic anomaly\cite{9302103,9309140}. The effect of this
is to add a non-holomorphic piece ${3 i\over 32\pi}\ln(S+\bar S)$
to $f(iS)$.
The modified
black hole entropy (see eqs.(4.12), (4.15) of
\cite{9906094}) reduces to, in the present case,
\be \label{eb1}
S_{BH} = 2 \pi \, {nw \over S_0+\bar S_0} 
- 12 \ln\left[ (S_0+\bar S_0) \, \eta\left( 
e^{-2\pi S_0}\right)^4\right]\, ,
\ee
where now $S_0$ is given by the solution of the equation
\be \label{eb2}
-{6\over \pi}\, \left[ 2 \, \p_{S_0} \ln\eta\left( 
e^{-2\pi S_0}\right) +{1\over S_0+\bar S_0}\right]
= {nw \over (S_0+\bar S_0)^2}\, .
\ee
The effect of holomorphic anomaly is represented by the term 
proportional to $\ln(S_0+\bar S_0)$ in \refb{eb1} and the term
proportional to $(S_0+\bar S_0)^{-1}$ in \refb{eb2}.
For real $S_0$, \refb{eb2} gives,
\be \label{eb4}
S_0 ={1\over 2}\, \sqrt{nw} + \, 
\OO\left(1\right)\, ,
\ee
and hence from \refb{eb1}
\be \label{eb6}
S_{BH} = 4\pi\sqrt{nw} - 12 \ln\sqrt{nw} + \OO(1)\, .
\ee

We can try to compare this with the logarithm of the degeneracy of
elementary string states. For a given $n$ and $w$ the degeneracy $d_{nw}$
is
determined by the formula\cite{rdabh1,rdabh2}:\footnote{Note 
the $N-1$ in the subscript
of $d$. This is due to the fact that for given $n$ and $w$, the required
level of left-moving oscillators is $N=nw+1$. Thus the associated
degeneracy is $d_{nw}=d_{N-1}$.}
\be \label{e7.9}
{16\over \Delta(q)} = q^{-1}\, 
\sum_{N=0}^\infty \, d_{N-1}\, q^{N}\, .
\ee
For large $N$, $d_N$ behaves as
\be \label{e7.10}
d_N \sim 8\, \sqrt 2\, N^{-27/4}\, \exp(4\pi\sqrt{N})\, .
\ee
Thus 
\be \label{e7.11}
S_{stat} = \ln (d_{nw}) \simeq 4\, \pi \, \sqrt{nw} - {27\over 2}
\, \ln \sqrt{nw} + \OO(1)\, .
\ee
Comparing \refb{e7.8a} with \refb{e7.11} we see that the quantum 
corrected Bekenstein-Hawking entropy does not correctly
reproduce the 
logarithmic
corrections to the statistical entropy. 

We should note
however that the definition of statistical
entropy itself can be ambiguous when we consider non-leading 
corrections.
Had we used the definition of statistical entropy based on
a different kind of ensemble instead of the 
microcanonical ensemble, we would
have gotten an answer that differs from \refb{e7.11}. Consider for example
the analog of the grand canonical ensemble where we introduce a 
chemical potential $\mu$ conjugate to $nw$ and introduce the 
partition function
\be \label{e9.2}
e^{\FF(\mu)} = \sum_{N=0}^\infty \, d_{N-1} \, e^{-\mu(N-1)}\, .
\ee
Then we can define the statistical
entropy through the thermodynamic relations 
\be \label{e9.3}
\wt S_{stat} = \FF(\mu) + \mu\, nw\, ,
\ee
where $\mu$ is obtained by solving the equation
\be \label{e9.4}
{\p \FF\over \p\mu} = -nw\, .
\ee
For large $nw$ we can approximate the sum in \refb{e9.2} as
$d_{N_0-1} e^{-\mu(N_0-1)}$ where $N_0$ is the value of $N$ that
maximizes the summand. \refb{e9.4} now
yields the result $N_0-1\simeq nw$
and \refb{e9.3} gives
$S_{stat}\simeq \ln d_{nw}$ to leading order. This
agrees with the definition of entropy based on the
microcanonical ensemble. However there are non-leading corrections
to this formula. To see this in the present context, note that
\refb{e7.9}, \refb{e9.2} give
\be \label{e9.5}
\FF(\mu) = \ln {16\over \Delta(e^{-\mu})} =
\ln {16\over \Delta(e^{-4\pi^2/\mu}) 
\left({\mu\over 2\pi}\right)^{-12}
} = {4\pi^2\over \mu} + 12\, \ln {\mu\over 2\pi}
+\ln 16 + \OO(e^{-4\pi^2/\mu})\, ,
\ee
where we have used the modular transformation law
of $\Delta(e^{-\mu})$ under
$\mu\to 4\pi^2/\mu$.
Eq.\refb{e9.4}, \refb{e9.3} then give,
\be \label{e9.6}
\mu = {2\pi\over \sqrt{nw}} +  
\OO\left({1\over \sqrt{nw}}\right)\, ,\qquad
\wt S_{stat} = 4\pi\, \sqrt{nw} - 12 \ln\sqrt{nw}+ \OO(1)
\, .
\ee
Thus we see that the logarithmic corrections present 
in \refb{e7.11} and \refb{e9.6} are different. In particular
\refb{e9.6} agrees with the black hole entropy given in
\refb{eb6}.

We can in fact do better and show that $\wt S_{stat}$ agrees with
$S_{BH}$ up to an additive constant of $\ln 16$ and
exponentially suppressed corrections. For this we
note that for $\FF(\mu)$ given in \refb{e9.5}, eqs.\refb{e9.3}, 
\refb{e9.4} take the form
\be \label{efirst}
\wt S_{stat} \simeq {4\pi^2\over \mu} + 12 \, \ln{\mu\over 2\pi}
+\ln 16 + \mu \, nw, \qquad -{4\pi^2\over \mu^2} +{12\over \mu}
\simeq -nw\, ,
\ee
where we have only ignored corrections of order $\exp(-4\pi^2/\mu)$.
On the other hand, ignoring terms of order $e^{-2\pi S_0}$, and
taking $S_0$ to be real,
eqs.\refb{eb1}, \refb{eb2} can be rewritten as
\be \label{esecond}
S_{BH} \simeq 4\, \pi \, S_0 - 12 \, \ln(2S_0) + {\pi n w \over S_0},
\qquad 4 S_0^2 - {12 S_0\over \pi} \simeq nw\, .
\ee
Comparing \refb{efirst} and \refb{esecond} we see that they are identical
up to an additive constant of $\ln 16$ in the 
expression for $\wt S_{stat}$,
if we make the identification
\be \label{ethird}
S_0={\pi\over \mu}\, .
\ee

We would like to add however that it
is not {\it a priori} obvious which definition of the
statistical entropy should be compared directly with the geometric
entropy. Thus comparison of the black hole and statistical entropy
beyond heterotic string tree level remains ambiguous.

\sectiono{Generalizations and Open Questions} \label{scomments}

\begin{enumerate}

\item {\bf Other four dimensional heterotic string theories:}
Instead of considering toroidal compactification of heterotic string
theory one could consider some other $N=2$ or $N=4$ compactification of
heterotic string theory for which the compact manifold has the form
$S^1\times K_5$ for some compact space $K_5$. It was argued in
\cite{9712150} that in all such cases if we consider a fundamental string
wrapped around the circle $S^1$ carrying $w$ units of winding and $n$
units of momentum along $S^1$, the entropy of the corresponding black hole
solution continues to be given by \refb{e20} with the same universal
constant $a$.  On the other hand the statistical entropy, computed from
the spectrum of the fundamental string wrapped on $S^1$,
is also given by the same formula \refb{e21} for large $n$ and $w$. Thus
once the agreement between \refb{e20} and \refb{e21} has been established
for the toroidally compactified heterotic string theory, it must continue
to hold for all other compactifications.

\item {\bf Higher dimensional heterotic string theories:}
It was shown in \cite{9506200} that the scaling argument of
\cite{9504147}, reviewed in section \ref{sreview}, continues
to hold for toroidally compactified heterotic string theory with higher
number of non-compact dimensions. Thus it is natural to ask if the
modification of the black hole solution, induced by the supersymmetric
generalization of the curvature squared
term in higher dimension, produces the
correct coefficient of the black hole entropy so that it agrees with the
statistical entropy obtained from computing the degeneracy of elementary
string states. At present the answer to this question is not known.

\item {\bf Other higher derivative corrections:}
In \cite{0409148,0410076} as well as in the present paper we have
taken into account only a specific class of higher derivative terms which
arise from supersymmetrization of the curvature squared
term. Since the
non-trivial modification of the solution takes place at $r\sim 1$ where
the higher derivative corrections are of order unity, there is no {\it a
priori} reason why further higher derivative corrections cannot completely
change the results. 
It will be interesing to explore if it is possible to analyze the
effect of these higher derivative terms by directly working with the
$\sigma$-model that describes string propagation in 
this background\cite{9408040}.

\item {\bf Type II string theories:}
One can carry out a similar analysis for toroidal or other 
compactification of type II
superstring theories which have at least two 
supersymmetries from the right-moving sector of the world-sheet and for 
which the compact space has a free circle on which one can wrap the 
fundamental string. It was shown in \cite{9712150} that the scaling 
argument of \cite{9504147} can be generalized to this case to yield a 
formula similar to \refb{e20}:
\be \label{e31} 
S_{BH} = a' \, \sqrt{nw}\, ,
\ee
where the constant $a'$ is universal for all superstring compactifications 
of the type mentioned above, but could differ from the constant $a$ for 
heterotic 
string compactifications. On the other hand the calculation of the 
degeneracy of the elementary string states yields the following expression 
for the statistical entropy:
\be \label{e32}
S_{stat} = 2\sqrt 2 \, \pi \sqrt{nw}\, .
\ee
Thus it is natural to ask if higher derivative corrections similar to the 
one studied here 
could give rise to the $a'=2\sqrt 2$ relation in superstring theory. 

Unfortunately however tree level
type II string theory has no curvature
squared term of the type discussed here and 
as a result the analog of the term that gave the correct value of $a$ 
in heterotic string compactification does not exist in type II string 
theory. Hence $a'$ continues to vanish. The resolution of this puzzle is 
not clear to us. It is of course possible that type II string theory will 
have other higher derivative corrections which modify the solution and 
gives us a finite entropy in agreement with \refb{e32}, but then the 
question that would arise is:
why are such corrections not present in the 
heterotic string theory?

Although we do not have an answer to this puzzle, the following
observation may be useful. First note that
the geometric entropy formula given in
\cite{9812082,0009234} always agrees with the apparent statistical
entropy computed in \cite{9711053}. Thus a discrepancy between the 
geometric entropy and statistical entropy can be regarded as a 
mismatch between the apparent statistical entropy computed in 
\cite{9711053} and the correct
statistical entropy, and understanding the origin of the latter
disagreement may give us some insight into the origin of the former
discrepancy. 
To this end we note that the analysis of \cite{9711053}
was carried out by describing the theory under 
consideration as an M-theory compactified on $S^1\times K_6$ for some
six dimensional Calabi-Yau manifold $K_6$, and the system whose
entropy is being computed as an M5-brane wrapped on 
$S^1\times K_4$ where
$K_4$ is a 4-cycle in $K_6$.
One then takes the limit of large
$S^1$ to regard this as a string wrapped on $S^1$, where the 
string is
identified as the M5-brane wrapped on $K_4$. The degeneracy of BPS
states, with the string carrying certain momentum along $S^1$, is 
then given by $\exp\left(2\pi\sqrt{c_L n/ 6}\right)$ 
where $c_L$ is the central charge associated
with the left moving modes on the string. Thus the statistical
entropy is given by $2\pi\sqrt{c_L n / 6}$.

The subtlety in this computation lies in the
determination of $c_L$. This requires
knowing the number of left-moving
massless modes living on the M5-brane wrapped
on $K_4$. In \cite{9711053} this computation was done by using certain 
genericity assumption under which the computation of the
number of massless degrees of 
freedom reduces to computation of certain topological
index. However in a
non-generic case the number of massless modes may differ from this index,
and in that case the entropy formula given in \cite{9711053} will not
be correct. Since the entropy formula of \cite{9711053}
is identical to the formula for the geometric entropy computed
in\cite{9812082,0009234}, this would imply that in these cases the
geometric entropy formula of \cite{9812082,0009234}
will differ from the statistical entropy.

\renewcommand{\wh}{\widehat}

Let us now examine the computation of \cite{9711053} both for the case
of heterotic string on $T^4\times \wt S^1\times S^1$ and type IIA 
on $T^4\times \wt S^1\times S^1$.
In the first case using string-string 
duality\cite{9410167,9501030,9503124,9504027,9504047}
we can map the
theory to type IIA on $K3\times \wt S^1\times S^1$, which in turn
is equivalent to M-theory on $\wh S^1\times K3\times \wt S^1
\times S^1$. Under this duality the fundamental heterotic string 
wrapped on $S^1$ gets mapped to M5-brane wrapped on $K3\times S^1$.
In this case the formula given in \cite{9711053} gives $c_L=c_2(K3)$
where $c_2(\MM)$ denotes the second Chern class of $\MM$. Since 
$c_2(K3)=24$,
we get the correct answer for the central charge associated with
the left-moving degrees of freedom of a fundamental heterotic string.
As a result geometric entropy agrees with the statistical entropy.

On the other hand using an analog of the string-string duality
formula we can map type IIA on $T^4\times \wt S^1\times S^1$ to
type IIA on $\wt T^4 \times \wt S^1\times S^1$ or M-theory on
$\wh  S^1\times \wt T^4 \times \wt S^1\times S^1$ 
so that the fundamental
type IIA string wrapped on $S^1$
 in the first theory gets mapped to the M5-brane 
wrapped on $\wt T^4\times S^1$\cite{9508064}. 
The formula given in \cite{9711053}
now gives $c_L=c_2(T^4)=0$. This clearly is not the correct answer
for the central charge of the left-moving modes on a type IIA string.
This
is responsible for the disagreement between the geometric entropy
formula of \cite{9812082,0009234} 
and the statistical entropy. 

This analysis shows that agreement between the geometric entropy
computed in \cite{9812082,0009234}
and the statistical entropy depends on whether the 
genericity assumption of 
\cite{9711053} holds or not. Thus in order to argue that the geometric
entropy always reproduces the statistical entropy, we need to show that
when the genericity assumption holds, there is a non-renormalization
theorem that prevents any correction to the entropy formula by higher
derivative terms which were not included in the analysis of
\cite{9812082,0009234}. On the other hand, when the genericity
assumption fails, the non-renormalization theorems must also
break down, and the higher derivative terms should become important.

\end{enumerate}

\medskip

{\bf Acknowledgement}: I wish to thank A.~Dabholkar, B.~de~Wit, R.~Gopakumar
and D.~Jatkar for useful 
discussions.

\end{document}